\documentclass[a4paper,11pt]{article}

\usepackage{graphicx}

\usepackage{epsfig,bm,amsmath}

\usepackage{jheppub} 

\usepackage[T1]{fontenc} 

\newcommand{\pa}{\mbox{\scriptsize\sf a}}    
\newcommand{\pb}{\mbox{\scriptsize\sf b}}    
\newcommand{\pq}{\mbox{\scriptsize\sf q}}    
\newcommand{\as}{\alpha_{\rm S}}                   
\newcommand{\ab}{\overline{\alpha}_{\rm S}}
\newcommand{\kk}{{\boldsymbol k}}            
\newcommand{\ku}{{\boldsymbol k}_1}
\newcommand{\kd}{{\boldsymbol k}_2}
\newcommand{\qq}{{\boldsymbol q}}
\newcommand{\dif}{{\rm d}}                   
\newcommand{\du}{\dif[\ku]}\newcommand{\dd}{\dif[\kd]}
\newcommand{\dk}{\dif[\kk]}

\newcommand{\G}{{\cal G}}                    
\newcommand{\K}{{\cal K}}                    
\newcommand{\e}{\varepsilon}                 
\renewcommand{\l}{\lambda}
\renewcommand{\o}{\omega}
\newcommand{\g}{\gamma}                   

\newcommand{\Eq}[1]{Eq.(\ref{#1})}         
\newcommand{\beq}{\begin{equation}}
\newcommand{\eeq}{\end{equation}}
\newcommand{\bea}{\begin{eqnarray}}
\newcommand{\eea}{\end{eqnarray}}
\newcommand{\nn}{\nonumber}

\title{Heavy quark impact factor in $k_T$-factorization}

\preprint{
\begin{flushright} 
IFIC/13-52 \\ LPN13-061 \\ \today
\end{flushright}\vspace*{-1cm}}

\author{Grigorios Chachamis,}
\author{Michal Deak}
\author{and Germ\'an Rodrigo}

\affiliation{
Instituto de F\'\i sica Corpuscular, \\ Universitat de Val\`encia -- 
Consejo Superior de Investigaciones Cient\'\i ficas,  \\
Parc Cient\'{\i}fic, E-46980 Paterna (Valencia), Spain.}

\emailAdd{grigorios.chachamis@ific.uv.es}
\emailAdd{michal.deak@ific.uv.es}
\emailAdd{german.rodrigo@csic.es}

\abstract{We present the calculation of the finite part of the heavy quark impact factor at 
next-to-leading logarithmic accuracy in a form suitable for phenomenological 
studies such as the calculation of the cross-section for
single bottom quark production at the LHC within the $k_T$-factorization scheme.}

\begin{document} 

\maketitle

\flushbottom

\section {Introduction}

One of the key issues in QCD phenomenology
at the Large Hadron Collider (LHC) is gauging the 
importance of small-$x$ physics related effects on
a number of physical observables and
consequently, getting a definite
answer on the validity and the applicability of the 
high energy resummation programme.

At very large center-of-mass energies, $\sqrt{s}$,
or alternatively at very small-$x$, the appearance of large
logarithms in energy ($\log s \sim  \log{1/x}$) 
can spoil the
convergence in the perturbative calculation of scattering amplitudes.
More concretely, terms of the form $(\as \, \log{1/x})^n$, 
where $\as$ is the strong coupling constant,  can be of order unity,
for small enough $x$, and
therefore need to be resummed to all orders. 
The Balitsky-Fadin-Kuraev-Lipatov (BFKL) 
framework enables  the resummation of high center-of-mass energy logarithms
at leading~\cite{BFKLLO, BFKLLO1, BFKLLO2} (L$x$)
and next-to-leading~\cite{BFKLNLO,BFKLNLO1} 
logarithmic accuracy (NL$x$). 
At L$x$, all the terms of the form  $(\as\, \log 1/x)^n$ are resummed whereas,
at NL$x$ one has also to resum terms in which the strong coupling lacks one power
compared to the logarithm in energy, 
that is, terms that behave like $\as\,(\as\, \log 1/x)^n$.

In the last two decades or so and
after the first small-$x$ data
from Deep Inelastic Scattering (DIS) collisions 
at HERA became available in the beginning of the 90's, the
resummation of the high energy logarithms 
and its phenomenological relevance exhibited some major developments.
To give but a sample of very important works in the field,
focusing more on the phenomenological side,
we would have to mention 
the formal study of the $k_T$-factorization 
scheme~\cite{Catani:1990xk, Catani:1990eg,Catani:1994sq}, the
computation of the NL$x$ BFKL kernel~\cite{BFKLNLO} and
the collinear improvements  to the NL$x$ kernel~\cite{Salam:1998tj, Ciafaloni:1998iv,
Ciafaloni:1999yw,Ciafaloni:2003rd}.
For the study of scattering amplitudes within the BFKL
formalism,
necessary ingredients are the gluon Green's function
which is obtained after solving the BFKL equation
(see for example Refs.~\cite{GGF, GGF1,GGF2,GGF3,GGF4,GGF5,GGF6,GGF7,
GGF8,GGF9,GGF10}), the gluon Regge trajectory~\cite{trajectory,trajectory1,
trajectory2,trajectory3,trajectory4,trajectory5,trajectory6,trajectory7,trajectory8,trajectory9} 
and
the impact factors~\cite{impact, impact1, impact2, impact3, impact4,
impact5, impact6, impact7, impact8}, the latter being process-dependent objects.
In a very general definition, 
the impact factors are the effective couplings of the scattering projectiles to 
whatever is exchanged in the $t$-channel for a process studied in
the $k_T$-factorization scheme.
One can claim that resumming small-$x$ logarithms is finally well
understood for a number of processes and
observables at HERA and the LHC:
perturbative evolution of parton distribution functions~\cite{Ball:1997vf,Ball:1997vf1,
Ball:1997vf2,Ball:1997vf3,Ball:1997vf4,Ball:1997vf5,Ball:1997vf6},
photoproduction~\cite{Catani:1990xk, Catani:1990eg,Catani:1994sq}
and double-DIS~\cite{Brodsky:1996sg,Brodsky:1996sg1,Brodsky:1996sg2} 
processes,
hadroproduction of heavy quarks~\cite{Ball:2001pq,Ball:2001pq1}, 
Drell-Yan~\cite{Marzani:2008uh},
Higgs boson hadroproduction~\cite{Hautmann:2002tu}, 
Mueller-Navelet jets and forward jets in DIS~\cite{Colferai:2010wu,Colferai:2010wu1,
Colferai:2010wu2,Colferai:2010wu3,Ducloue:2013bva}.

The impact factors for gluons and  massless quarks were
calculated in Ref.~\cite{Ciafaloni:1998hu} at NL$x$ accuracy
and in momentum space.
This allows in principle for the calculation of various DIS and double-DIS
processes with massless quarks and gluons in the initial state whereas
the extension to the case of hadron-hadron collisions was also  
established~\cite{Mueller:1986ey,Vera:2007kn,Kwiecinski:2001nh}.

What we are interested in though,
starting from this work, is to set up a programme
for phenomenological studies --within the $k_T$-factorization scheme--
of processes involving massive quarks (mainly bottom quarks) at the LHC,
given the excellent tagging capabilities of the ATLAS~\cite{ATLAS:2011qia},
 CMS~\cite{Chatrchyan:2012jua} and LHCb~\cite{LHCb:2003ab} detectors.
For that to be possible we need 
the NL$x$ impact factor for a massive quark which was first 
calculated by Ciafaloni and Rodrigo 
in Refs.~\cite{Ciafaloni:2000sq,Ciafaloni:2000sq1}. 
However, their final expression for the massive quark impact factor was
written in the form that contains a sum of an infinite number of terms. 
To make their result directly applicable for
phenomenological studies we recalculate the NL$x$ heavy quark 
impact factor in a compact and resummed form, ready to be used
for the convolution with the 
gluon Green's function in a numerically
straightforward manner. 

After this introduction, we proceed to Section 2 where
we set up our notation and provide the necessary definitions. In Section 3,
we present the two terms that contribute with massive corrections to the massless
NL$x$ quark impact factor and we calculate these terms in Sections 4 and 5.
The full result in a closed resummed form appears in Section 6,
in which we also offer a first numerical study of the behavior of the finite part of the result.
Finally, we conclude in Section 7.

\section{High energy factorization}

In the high energy limit: $\Lambda_{QCD}\ll |t|\ll s$, the 
partonic cross-section of 
$2\rightarrow 2$ processes factorizes into the impact factors $h_{\pa}(\ku)$ and 
$h_{\pb}(\kd)$ of the two colliding partons ${\large\sf a}$ and ${\large\sf b}$,
and the gluon Green's function $\G_{\o}(\ku,\kd)$ (here in Mellin space) so
that the differential cross-section can be written as
\beq
 \frac{\dif\sigma_{\pa\pb}}{\du\,\dd}=
 \int\frac{\dif\o}{2\pi i\, \o}\, h_{\pa}(\ku) \, \G_{\o}(\ku,\kd) \,
 h_{\pb}(\kd)\left(\frac{s}{s_0(\ku,\kd)}\right)^{\o}~,
\label{fatt}
\eeq
where $\o$ is the dual variable to the rapidity Y,  and
$\dk=\dif^{2+2\e}\kk/\pi^{1+\e}$ is the transverse space measure.
The impact factor at the leading $\log x$ order (L$x$)
can be expressed by a very simple formula
\beq
 h_{}^{(0)}({\kk})=\sqrt{\frac{\pi}{N_c^2-1}}\;
 \frac{2C_F \as N_{\e}}{\kk^2\,\mu^{2\e}}~, \quad
 N_{\e} = \frac{(4\pi)^{\e/2}}{\Gamma(1-\e)}~,
\label{hzero}
\eeq
and it is the same (up to a color factor)
for quarks and gluons, where
$\mu$ is the renormalization scale and $\e$ is the dimensional
regularization parameter. Using the expression for the leading order 
impact factor we define the constant 
\beq\label{eq:Aep}
A_{\e} =  \kk^2 \, h_{}^{(0)}({\kk}) \,
 \frac{\ab}{\Gamma(1-\e)\mu^{2\e}}~,
\eeq
which contains the dependence on the strong coupling
and on color factors.
The dimensionless strong coupling $\ab$ is expressed by using the gauge
coupling parameter $g$ and already introduced parameters:
\beq\label{eq:alphas}
\ab = \frac{\as N_c}{\pi}~, \qquad
\as = \frac{g^2\Gamma(1-\e)\mu^{2\e}}{(4\pi)^{1+\e}}~,
\eeq
where $N_c$ is the number of colors in QCD.
Finally, the gluon Regge trajectory, $\o^{(1)}(\kk)$, which
accounts for the virtual correction to the BFKL kernel, has the simple form:
\beq\label{eq:omega}
 \o^{(1)}(\kk)=-\frac{g^2 N_c \kk^2}{(4\pi)^{2+\e}} \,
 \int\frac{\dif[{\boldsymbol p}]}{{\boldsymbol p}^2(\kk-{\boldsymbol p})^2}=
 -\frac{\ab}{2\e}\frac{\Gamma^2(1+\e)}{\Gamma(1+2\e)}
 \left(\frac{\kk^2}{\mu^2}\right)^{\e}~.
\eeq
\section{The integral representation of the impact factor}

According to Ref.~\cite{Ciafaloni:2000sq},
the NL$x$ result for the impact factor 
of a heavy quark can be written as the sum of three contributions:
\begin{align}
h_{\pq}^{(1)}(\kd) &= h_{\pq,m=0}^{(1)}(\kd) 
+ \int_0^1 \dif z_1 \int \du \Delta F_{\pq}(z_1,\ku,\kd) \nn \\
& + \int \du \, \ab \, h_{\pq}^{(0)}(\ku) \, K_0(\ku,\kd) \,
  \log \frac{m}{k_1} \, \Theta_{m \, k_1}~,
\label{eq:h1mass}
\end{align}
with the convention 
$\Theta_{m \, k_1}=\theta\left(m - k_1\right)$, where the 
 $\theta$-function is the well-known Heaviside step function
 and $k_1 = |\ku|$.
The first term on the right hand side of~\Eq{eq:h1mass} is the NL$x$
correction to the impact factor of a massless quark, 
which can be expressed by using the leading
order impact factor $h_{}^{(0)}({\kk})$ in~\Eq{hzero} and the gluon Regge 
trajectory $\o^{(1)}(\kk)$ in~\Eq{eq:omega},
\beq
h_{\pq,m=0}^{(1)}(\kd)
= h^{(0)}(\kd) \, \o^{(1)}(\kd) \, \left[ b_0
+ \frac{3}{2} - \e \, \left( \frac{1}{2} + \K  \right) \right]~,
\label{eq:h1massless0}
\eeq
with the beta function $b_0$ and $\K$ defined as
\beq
b_0 = \frac{11}{6} - \frac{n_f}{3N_c}~, \qquad
\K = \frac{67}{18}-\frac{\pi^2}{6}-\frac{5n_f}{9N_c}~.
\eeq

The second term on the right hand side of~\Eq{eq:h1mass}
is the NL$x$ correction induced
by the heavy quark mass $m$, with $\Delta F_{\pq}(z_1,\ku,\kd)$ defined 
in Ref.~\cite{Ciafaloni:2000sq}. 
The third term comes from the introduction of 
the mass scale to the leading order BFKL kernel $K_0(\ku,\kd)$ 
which is defined as
\beq
\label{eq_k0}
\ab \, K_0(\ku,\kd) = 
\frac{\ab}{\qq^2 \Gamma(1-\e)\mu^{2\e}}
+ 2 \o^{(1)}(\ku) \delta[\qq]~, \qquad 
\delta[\qq]=\pi^{1+\e} \delta^{2+2\e}(\qq)~,
\eeq
with $\qq=\ku+\kd$.
The first term on the right hand side of~\Eq{eq_k0} is the real 
component of the BFKL kernel
and the second one corresponds to the virtual corrections.
In the following two Sections we reanalyze the second and third terms in
the right hand side 
of~\Eq{eq:h1mass}.

\section{The $\Delta F_{\pq}$ term}

The second term in the right hand side of~\Eq{eq:h1mass}, 
\beq
\Delta F_{\pq}(\kd) =
\int_0^1 \dif z_1 \int \du \Delta F_{\pq}(z_1,\ku,\kd)~,
\label{eq:deltaF2}
\eeq
receives contributions from virtual and real corrections. 
The explicit expression of the integrand of~\Eq{eq:deltaF2}
is given in Ref.~\cite{Ciafaloni:2000sq} in momentum space, 
after integration over $\ku$.
Note, however, that the remaining integrations cannot be performed 
directly in an straightforward way. Instead, it is easier to 
calculate the Mellin transform:
\beq\label{eq:impfnlo}
\Delta \tilde{F}_{\pq}(\gamma) =  
\Gamma(1+\e) \, (m^2)^{-\e} \int \dd  \left( \frac{\kd^2}{m^2}\right)^{\g-1}
\Delta F_{\pq}(\kd)~,  
\eeq
which leads to this expression
\begin{align}\label{eq:Melltr}
\Delta\tilde{F}_{\pq}(\gamma) &= A_{\e} \, (m^2)^{\e} \,
\frac{\Gamma(\g+\e)\Gamma(1-\g-2\e)\Gamma^2(1-\g-\e)}
{8\Gamma(2-2\g-2\e)} \nn \\ & \times 
\bigg[ \frac{1+\e}{\g+2\e} + \frac{2}{1-2\g-4\e}
\left( \frac{1}{1-\g-2\e}- \frac{1}{3-2\g-2\e} \right) \bigg]~.
\end{align}

Then, the function $\Delta F_{\pq}(\kd)$ in momentum space is recovered 
by computing the inverse Mellin transform: 
\beq\label{eq:invMell}
\Delta F_{\pq}(\kd) =
\frac{1}{m^2} \int_{1-2\e < {\rm Re}\, \g <1-\e} \frac{\dif \g}{2\pi i}  
\left(\frac{\kd^2}{m^2}\right)^{-\g-\e}
\Delta \tilde{F}_{\pq}(\gamma)~.  
\eeq
This integral is a contour integral in the complex plane which is well
defined when the integration contour is a straight line
parallel to the imaginary axis and which intersects the real
axis in the strip $1-2\e < {\rm Re}\, \g <1-\e$.
To perform the integration in~\Eq{eq:invMell} we use Cauchy's residue
theorem.
The ratio $\kd^2/m^2$ may, in principal, take any value between $0$
and $\infty$. If $\kd^2/m^2 < 1$, we deform the integration
contour at $- i \infty$ and at $+ i \infty$ to the right, such that the two ends meet at
$+ \infty$  of the real axis, whereas, if  $\kd^2/m^2 > 1$
we deform the integration
contour at $- i \infty$ and at $+ i \infty$ to the left, such that the two ends meet at
$- \infty$  of the real axis. In both cases, we change the initial integration contour
to a closed one which consists of the initial one and of a semi-circle on which the integrand
is zero, allowing this way for the integration to be done by summing the residue
contributions enclosed by each contour. It turns out though, that in order to obtain
the result in a simple resummed form, 
we are forced to close the contour to the left,
assuming that $\kd^2/m^2 > 1$, in which case we denote the result by 
$\Delta F_{\pq}^-(\kd)$.
We have checked that this closed resummed result is the correct result
for all allowed values of the ratio $\kd^2/m^2$
by comparing to the expression --denoted as $\Delta F_{\pq}^+(\kd)$--
we get after closing the initial contour to the right.
It is probably noteworthy to add that in the case of deforming the contour
to the left, we were able to resum the residue
contributions in a closed form both before and after expanding in $\e$
whereas in the case of closing the contour to the right, the resummation
of the residue contributions in a closed form is only possible before
expanding in $\e$.

In detail, after deforming the integration contour as described above, we have:
\beq
\Delta F_{\pq}^-(\kd) = \frac{1}{m^2} \, \sum_{\gamma\le 1-2\e} {\rm Res} \, 
\left[ \left( \frac{\kd^2}{m^2}\right)^{-\g-\e}
\, \Delta\tilde{F}_{\pq}(\gamma) \right]~.
\eeq

The distinct pole contributions 
that need to be accounted for come from the residues at $\gamma = 1- 2 \e$
(which provides the singular terms in $\e$),
$\frac{1}{2} - 2 \e$, $ - \e$, $ - 2 \e$ and finally from the poles at
$\gamma = -n - \e$ with $n\in \mathbb{N}$ and $n>0$. 
To simplify the formalism, we factorize the leading order impact 
factor $h_{}^{(0)}({\kd})$ and the gluon Regge trajectory $\o^{(1)}(\kd)$
at each residue contribution and define 
\beq
h_{\gamma_i}(\kd) = \left(h_{}^{(0)}({\kd}) \, \o^{(1)}(\kd)\right)^{-1} \, 
\frac{1}{m^2} \, {\rm Res}_{\{\gamma = \gamma_i\}}
\left[ \left( \frac{\kd^2}{m^2}\right)^{-\g-\e}
\, \Delta\tilde{F}_{\pq}(\gamma) \right]~.
\eeq
Then
\beq
\Delta F_{\pq}^-(\kd) = h^{(0)}({\kd}) \, \o^{(1)}(\kd) \,  
\sum_{\gamma_i \le 1-2\e} h_{\gamma_i}(\kd)~
\eeq
and the contributions of the different residua at the poles 
located at $\gamma \le 1-2\e$ are given by
\bea
\label{eq:c5} 
h_{1-2\e}(\kd) &=& - \frac{1+5\e-2\e^2}{2(1+2\e)} - \log(4R) + 
\psi(1-\e) - \psi(1)- 2 \psi(\e) + 2 \psi(2\e)~,  \nn \\
h_{1/2-2\e}(\kd) &=&  \sqrt{R} \, 
\frac{(3+4\e) \, \Gamma(1+2\e) \, \pi \tan(\pi\, \e)}{4^{1+\e} (1+\e) \, \Gamma^2(1+\e)}~, \nn \\
h_{-2\e}(\kd) &=& R \, \frac{1+\e}{1+2\e}~, \nn \\
h_{-n-\e}(\kd) &=&  (-1)^{1+n}\, (4R)^{1+n-\e} \, 
\frac{\e \,\Gamma(1+2\e) \Gamma(1+n)\, \Gamma(1+n-\e)}{4 \, \Gamma(1-\e)\, \Gamma^2(1+\e) \, \Gamma(2+2n)} \, \nn \\ &\times&
\left[ \frac{1+\e}{\e-n} + \frac{2}{1+2n-2\e} \left(\frac{1}{1+n-\e}-\frac{1}{3+2n} \right)
\right]~, 
\eea
where $R=\kd^2/(4m^2)$. The residue at $\gamma=-\e$ is accounted for when $n=0$
in the last expression in \Eq{eq:c5}. 
To obtain a closed analytic expression we
still need to resum  $h_{-n-\e}(\kd) $ for $n \ge 1$.
The result is given by:
\bea 
\sum_{n=1}^{\infty} h_{-n-\e}(\kd) &=& - (4R)^{2-\e} \frac{\e \, \Gamma(1+2\e)}{24 \, \Gamma^2(1+\e)}
\bigg[ (1+\e) \, {}_2F_1(1,1-\e;\frac{5}{2};-R) \nn \\ 
&-& \frac{4(1-\e)}{3-2\e} \, {}_3F_2(1, \frac{3}{2} - \e, 2 - \e; \frac{5}{2}, \frac{5}{2} - \e; -R) \nn \\ 
&+& \frac{2(1-\e)}{5(3-2\e)} \, {}_3F_2(1, \frac{3}{2} - \e, 2 - \e; \frac{7}{2}, \frac{5}{2} - \e; -R) \nn \\
&+& \frac{2(1-\e)}{2-\e} \, {}_3F_2(1, 2 - \e, 2-\e; \frac{5}{2},3-\e; -R) \bigg]~,
\eea 
where $_{p}F_{q}$ are generalized hypergeometric functions.
By expanding in $\e$ we obtain
\bea \label{eq:sumc5}
&& \sum_{n=1}^{\infty} h_{-n-\e}(\kd) = 
\e \, \Bigg[ 1 + \frac{10}{3} \, R
+ \log(Z) \left( (1+2R) \sqrt{\frac{1+R}{R}} + 2 \log(Z) \right) \nn \\ 
&& + 3 \sqrt{R} \, \left( {\rm Li}_2(Z) - {\rm Li}_2(-Z) 
+ \log(Z)  \, \log \left( \frac{1-Z}{1+Z} \right) - \frac{\pi^2}{4} \right) \Bigg] 
+ {\cal O}(\e^2)~, 
\eea
where $\text{Li}_2$ is the usual dilogarithm function, 
and $Z=(\sqrt{1+R}+\sqrt{R})^{-1}$. 
The final result is obtained by summing up the contributions of all 
the residua and reads
\bea
\Delta F_{\pq}^⁻(\kd) &=& h^{(0)}({\kd}) \, \o^{(1)}(\kd) \, \Bigg\{
\frac{1}{\e} - \log(4R) -\frac{1}{2} \nn \\ &+& \e \, \Bigg[ \frac{\pi^2}{6} 
- \frac{1}{2} + R \, \log(4R)
+ \log(Z) \left( (1+2R) \sqrt{\frac{1+R}{R}} + 2 \log(Z) \right) \nn \\ 
&+& 3 \sqrt{R} \, \left( {\rm Li}_2(Z) - {\rm Li}_2(-Z) 
+ \log(Z)  \, \log \left( \frac{1-Z}{1+Z} \right) \right) \Bigg]
\Bigg\} + {\cal O}(\e)~. 
\label{eq:deltaFq}
\eea

What we have achieved so far is a compact analytic expression for 
the second term in~\Eq{eq:h1mass}, that is, $\Delta F_{\pq}(\kd)$, 
for any value of $R$ keeping the whole singularity structure.
The double log singularities ($1/\e^2$ and $1/\e ~\log(R)$ terms)
cancel~\cite{Ciafaloni:2000sq} against the double log singularities 
from the remaining third term of~\Eq{eq:h1mass}.

\section{The $K_0(\ku,\kd)$ related term}

Let us now turn to the final ingredient in order to have the full NL$x$ heavy
quark impact factor with mass corrections.
For the real emission part of the BFKL kernel, $K_0(\ku,\kd)$ 
(see~\Eq{eq_k0}), we define the integral
\beq
  I_m(\kd) = \int \du \frac{\ab  h_{\pq}^{(0)}({\ku})}{\qq^2 \Gamma(1-\e)\mu^{2\e}}
  \log \frac{m}{k_1} \, \Theta_{m \, k_1}~. 
\eeq
We use the following integral representation~\cite{Ciafaloni:2000sq}:
\beq\label{eq:logrep}
 \log\frac{a}{b} \,\Theta_{ab}=\lim_{\alpha\to0^+}
 \int_{-i\infty}^{+i\infty}\frac{\dif\l}{2\pi i}\,\frac{1}{(\l+\alpha)^2}
 \left(\frac{a}{b}\right)^{\l}
 \equiv \int \dif[\l] \, \left(\frac{a}{b}\right)^{\l}~,
\eeq
valid for $a,b>0$, which allows us to write
\begin{align}
I_m(\kd) & = \frac{A_{\e}}{2}
\int \dif[\l] \, (m^2)^{\l} \int \frac{\du}
{\qq^2 \, (\ku^2)^{1+\l}}  \nn \\
& = \frac{A_{\e}}{2} \int \dif[\l]\, 
\frac{\Gamma(1+\l-\e) \Gamma(\e) \Gamma(\e-\l)}
{\Gamma(1+\l) \Gamma(2\e-\l)} (m^2)^{\l} (\kd^2)^{-1-\l+\e}~,
\label{Im}
\end{align}
or more explicitly
\beq
I_m(\kd) =
 \frac{A_{\e}}{2} \, \lim_{\alpha\to0^+} \, \int_{-i \infty}^{+i  \infty} 
 \frac{d\l}{2 \pi i} \, \frac{1}{(\l + \alpha)^2}
\frac{\Gamma(1+\l-\e) \Gamma(\e) \Gamma(\e-\l)}
{\Gamma(1+\l) \Gamma(2\e-\l)} (m^2)^{\l} (\kd^2)^{-1-\l+\e}~.
\label{Imdetail}
\eeq
The integrand in~\Eq{Imdetail} 
vanishes for $|\l|\rightarrow \infty$ in all directions
apart from the real axis.
As was the case in the previous Section, this is a contour integral and
therefore we define
\bea
&& d_{\l_i}(\kd) = \frac{A_{\e}}{2} \,  
\left(h_{}^{(0)}({\kd}) \, \o^{(1)}(\kd)\right)^{-1} \, \nn \\ && \qquad \times
\lim_{\alpha\to 0^+} \, {\rm Res}_{\{\l = \l_i\}} \, \left[\frac{1}{(\l + \alpha)^2}
\frac{\Gamma(1+\l-\e) \Gamma(\e) \Gamma(\e-\l)}
{\Gamma(1+\l) \Gamma(2\e-\l)} (m^2)^{\l} (\kd^2)^{-1-\l+\e}\right]~.
\label{dImdetail}
\eea

For $m^2/\kd^2 < 1$ we close the contour at infinity to the right of
the complex plane, and evaluate the residua of the poles enclosed 
by the deformed contour. The first pole for $\l>0$ is located at $\l=\e$ which gives 
the leading contribution including a singular term in $\e$. 
All remaining poles are located at $\l = n+\e$ with $n\in \mathbb{N}$
and give contributions of order $(m^2/\kd^2)^n$. 
The complete expression for the residua of these poles,
including the one at $n=0$, reads:
\beq\label{eq:resdp2}
d_{n+\e} (\kd) = 
\frac{\Gamma(1+2\e)}{\Gamma(1-\e)\, \Gamma(1+\e)} \, 
\frac{(-1)^n \, (4R)^{-n-\e}}{(\e+n)^3 \, \Gamma(\e-n) \, \Gamma(\e+n)}~.
\eeq 
Keeping apart the contribution of the first pole at $n=0$, we resum the 
series of residua at $\l = n+\e$ with $n\ge 1$ and  we obtain
\bea\label{eq:sumdp2}
\sum_{n=1}^{\infty} d_{n+\e} (\kd) &=&
\frac{\e \, (1-\e) \, \Gamma(1+2\e)}{\Gamma(1-\e)\, \Gamma^3(2+\e)}  
\nn \\ &\times&
\frac{1}{(4R)^{1+\e}} \, {}_4F_3\left(1,2-\e ,1+\e ,1+\e ;2+\e ,2+\e ,2+\e;\frac{1}{4R}\right)\,.
\eea
By expanding in $\e$ the contribution of the 
first pole of~\Eq{eq:resdp2} at $n=0$ and the result in~\Eq{eq:sumdp2} and after
summing them up we get:
\bea
I_m^+(\kd) &=& - h^{(0)}({\kd})\, \o^{(1)}(\kd) \, \sum_{\l_i > 0} d_{\l_i} (\kd)  \nn \\
&=& h^{(0)}({\kd})\, \o^{(1)}(\kd) \,
\left[-\frac{1}{\e} + \log(4R) 
- \e\, \left(\frac{1}{2} \log^2(4R) + 
{\rm Li}_2\left(\frac{1}{4R}\right)\right)\right] + {\cal O}(\e)~.
\label{eq:Implus}
\eea

We need to consider in addition the case
$\kd^2/m^2 < 1$. Now, we close the integration contour to the left,
to $- \infty$, enclosing this way the poles located 
at $\l = - \alpha$ with $\alpha \to 0^+$
and at $\l = -n +\e$, with $n \in \mathbb{N}$ and $n \ge 1$, with
\beq
I_m^-(\kd) = h^{(0)}({\kd})\, \o^{(1)}(\kd) \, \sum_{\l_i < 0} d_{\l_i} (\kd)~.
\eeq
These residua provide contributions of the order $(\kd^2/m^2)^n$,
their actual values are:
\bea \label{eq:resd1}
\left. d_{-\alpha} (\kd)\right|_{\alpha \to 0^+} &=&
2 \log\left(4R\right) + 2 \left[ \psi(1) - \psi(1-\e ) + \psi(\e ) - \psi(2\e) \right]~,  \\
\label{eq:resd2}
d_{-n+\e}(\kd) &=& \frac{\Gamma(1+2\e)}{\Gamma(1-\e)\, \Gamma(1+\e)} \, 
\frac{(-1)^n \, (4R)^{n-\e}}{(\e-n)^3 \, \Gamma(\e-n) \, \Gamma(\e+n)}~.
\eea
As before, we resum all the residua at $\gamma=-n +\e$ 
before we expand in $\e$. The result is:
\bea\label{eq:sumd2}
\sum_{n=1}^{\infty} d_{-n+\e} (\kd) &=& 
- \frac{\Gamma(1+2\e) \, \sin(\pi\e)}{\pi\, (1-\e)^2 \, \Gamma(1+\e)^2} \nn \\ &\times& 
(4R)^{1-\e} \, {}_4F_3\left(1,1-\e,1-\e ,1-\e;2-\e ,2-\e ,1+\e; 4R\right)~.
\eea
After summing up the contributions from~\Eq{eq:resd1} 
and~\Eq{eq:sumd2}, expanding in $\e$, and including 
the virtual term of the BFKL kernel $K_0(\ku,\kd)$, we obtain:
\bea
I_m^-(\kd) &-& h^{(0)}({\kd})\, \o^{(1)}(\kd) \, \log \left(\frac{\kd^2}{m^2}\right) 
\nn \\ &&
= h^{(0)}({\kd}) \, \o^{(1)}(\kd) \,
\left(-\frac{1}{\e}+\log(4R)-\e\,{\rm Li}_2(4R)\right) + {\cal O}(\e)~.
\label{eq:Imminus}
\eea

\section{The analytic result for the impact factor and a first numerical study}

The final expression for the next-to-leading order correction $h_{\pq}^{(1)}(\kd)$ 
to the impact factor is obtained by using \Eq{eq:h1massless0},
\Eq{eq:deltaFq}, \Eq{eq:Implus} and \Eq{eq:Imminus} into \Eq{eq:h1mass}.
Collecting all the contributions, the impact factor of a heavy quark 
at NL$x$ accuracy reads 
\beq
h_{\pq}(\kd) = h^{(0)}(\kd) + h_{\pq}^{(1)}(\kd) 
\eeq
and can be expressed in terms of a singular and a finite contribution
\beq\label{eq:h(1)}
h_{\pq}(\kd) = h_{\pq}^{(1)}(\kd)|_{\rm sing} + h_{\pq}(\kd)|_{\rm finite}~.
\eeq
The singular term $h_{\pq}^{(1)}(\kd)|_{\rm sing}$ reads~\cite{Ciafaloni:2000sq} 
\beq
h_{\pq}^{(1)}(\kd)|_{\rm sing} =
h^{(0)}(\kd)  \bigg(
\frac{3}{2}  \, \o^{(1)}(\kd)
-\frac{1}{2} \, \o^{(1)}(m) \, \Theta_{k_2 \, m}
-\frac{1}{2} \, \o^{(1)}(\kd) \, \Theta_{m \, k_2} \bigg)~,
\eeq
whereas the finite contribution, which is the main result of this paper, 
is given by 
\beq
\begin{split}
h_{\pq}(\kd)|_{\rm finite} & =
h^{(0)}(\kd, \as(\kd)) \, \Bigg\{1 + \frac{\as \, N_c}{2\pi} \, \Bigg[ 
\mathcal{K}-\frac{\pi^2}{6} + 1 - R \, \log(4R) \\ &   
- \log(Z) \bigg(\left(1+2R\right)\sqrt{\frac{1+R}{R}} + 2 \log(Z) \bigg) \\ & 
-  3 \, \sqrt{R} \left({\rm Li}_2(Z)-{\rm Li}_2(-Z)
+ \log(Z) \, \log \left(\frac{1-Z}{1+Z} \right) \right) \\
& + {\rm Li}_2\left(4R\right) \, \Theta_{m\, k_2} + \left(
  \frac{1}{2} \log\left(4R\right)+\frac{1}{2} \log^2\left(4R\right)
+ {\rm Li}_2 \left( \frac{1}{4R} \right) \right) \, \Theta_{k_2\,m}  
\Bigg] \Bigg\}~.
\label{eq:finalfinite}
\end{split}
\eeq
As in Ref.~\cite{Ciafaloni:2000sq}, we have absorbed the singularities 
proportional to the beta function $b_0$ into the running 
of the strong coupling $\as(\kd)$~\cite{Rodrigo:1993hc,Rodrigo:1997zd}.
Our final result in~\Eq{eq:finalfinite} is valid in any kinematical 
regime, and provides a compact expression for the heavy quark 
impact factor which is suitable for phenomenological studies. 
We have checked that by expanding~\Eq{eq:finalfinite}
for either $\kd^2/m^2 < 1$ or $m^2/\kd^2 < 1$ 
we reproduce the results presented in Ref.~\cite{Ciafaloni:2000sq}\footnote{
There were two typos in Ref.~\cite{Ciafaloni:2000sq} which were taken
into account when we made the comparison.}.
In particular, the massless limit of the finite 
contribution to the impact factor in~\Eq{eq:finalfinite}
is given by
\beq
h_{\pq}(\kd,m=0)|_{\rm finite}  =
h^{(0)}(\kd, \as(\kd)) \, \left\{ 1 + \frac{\as \, N_c}{2\pi} \, \left[ 
\mathcal{K}-\frac{\pi^2}{6} - \frac{3}{2} \right] \right\}~.
\eeq

With the result from~\Eq{eq:finalfinite} at hand, we proceed
to a first numerical study of the magnitude of the mass corrections
to the impact factor at NL$x$ accuracy.
As was stated previously, 
we have adopted the running coupling scheme
as described in Refs.~\cite{Rodrigo:1993hc,Rodrigo:1997zd} 
and with $n_f=5$ flavors. At L$x$ accuracy,  we use
a fixed value for the strong coupling constant, namely, $~\ab = 0.2$.
\vspace{2cm}
\begin{figure}[tbh]
\vspace{0cm}
  \begin{picture}(0,0)
    \put(30, -220){
      \includegraphics{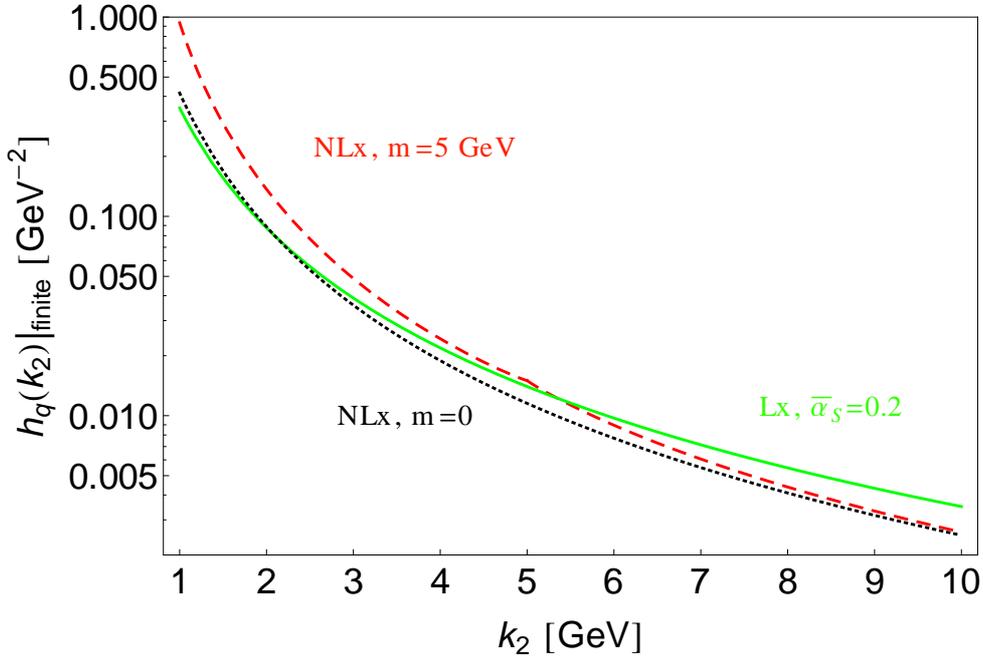}
    }    
     \end{picture}
\vspace{7.8cm}
\caption{Finite part of the quark impact factor: L$x$ in green (solid line), 
massless NL$x$ in black (dotted line) and
NL$x$ for quark mass $m=5$~GeV in red (dashed line).}
\label{fig:impfs}
\end{figure}

\begin{figure}[tbh]
\vspace{1cm}
  \begin{picture}(0,0)
    \put(27, -210){
      \includegraphics{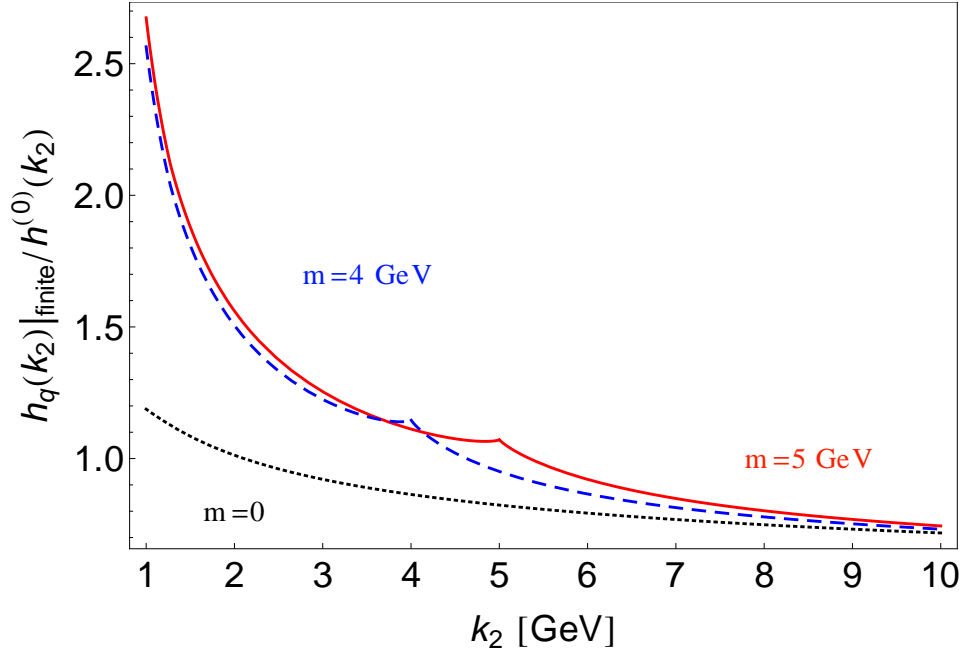}
    }    
     \end{picture}
\vspace{7.5cm}
\caption{The ratio of the NL$x$ impact factor to the L$x$ impact factor for 
different quark masses: $m=0$~GeV (massless) in black (dotted line), 
$m=4$~GeV in blue (dashed line)
and $m=5$~GeV in red (solid line).}
\label{fig:rimpfs}
\end{figure}

\begin{figure}[tbh]
\vspace{1cm}
  \begin{picture}(30,0)
    \put(27, -210){
      \includegraphics{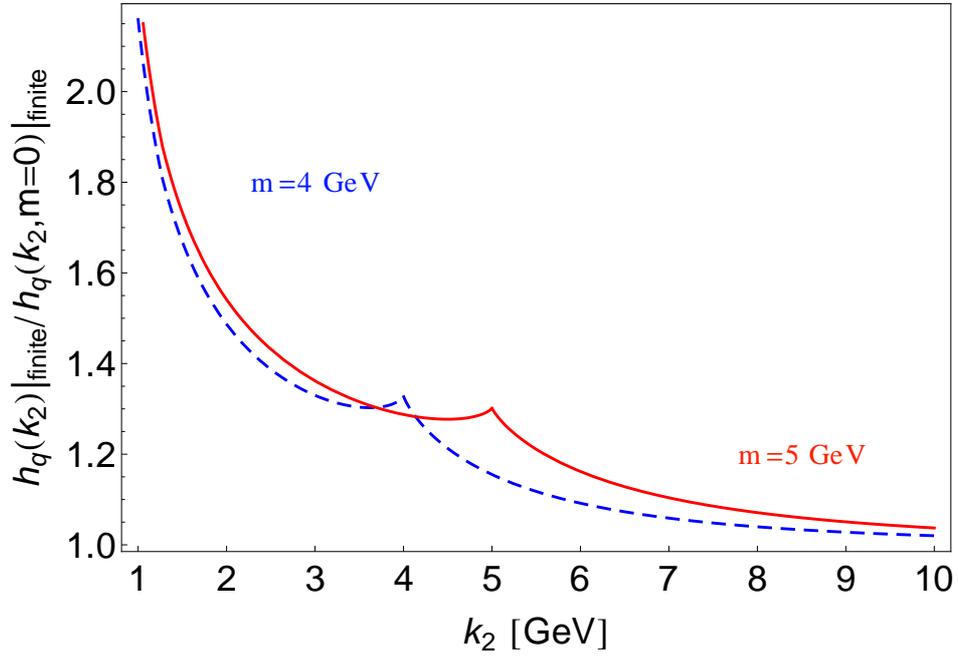}
    }    
     \end{picture}
\vspace{7.5cm}
\caption{Ratio of the massive NL$x$ impact factor over the massless NL$x$ factor 
for two different mass choices: $m=4$~GeV in blue (dashed line) 
and $m=5$~GeV in red (solid line).
}\label{fig:rimpfsmm}
\end{figure}

In Fig.~\ref{fig:impfs} we plot the L$x$ as well as the
NL$x$ quark impact factor, the latter for two quark mass choices, 
$m=0$ and $m=5$~GeV. We see that the NL$x$
correction to the leading order impact
factor for massless quark is positive and moderate only for
small $\kd$ where the behavior is dominated by the running of the 
strong coupling constant, whereas, for most of the range of the plot the correction
 is negative.
For a non-zero quark mass
the overall correction is positive and large in the region
$\kd^2/m^2 < 1$. They turn to negative closely
after $\kd^2/m^2  = 1$ and for larger  $\kd^2/m^2$,
they follow the NL$x$ massless curve as expected.
To get a better quantitative picture of the behavior 
of the NL$x$ corrections in the 
massless and massive case, 
it is useful to study the ratios of the
impact factors at L$x$ and NL$x$ accuracy.
In Fig.~\ref{fig:rimpfs} we can see that the relative size of the
full NL$x$ corrections in the range
$\kd^2/m^2 < 10$ GeV, vary from more than $+100\%$ at very small
$\kd$ down to some $-20\%$ for larger $\kd$, for 
similar mass choices $m=4$ GeV and $m=5$ GeV.
In Fig.~\ref{fig:rimpfsmm} the ratio between the finite parts of the NL$x$ 
massive and massless quark impact factor is plotted. The corrections 
induced purely by a non-zero mass are of the order of a $100\%$ 
in the small $\kd^2/m^2$ limit and decrease as $\kd$ is getting larger. 
The cusps in the curves are solely an artefact of the choice of the 
factorization scale, for details we refer the reader to Section 3 of Ref.~\cite{Ciafaloni:2000sq}.
As expected, in the limit $\kd\rightarrow\infty$, the massless and massive 
NL$x$ impact factors coincide such that their ratio approaches $1$.

\section{Conclusions}

In this work, we re-calculated 
the heavy quark impact factor 
at next-to-leading logarithmic accuracy 
and we obtained a closed analytic
result for its finite part,
suitable for an immediate numerical implementation.
We performed a first comparative numerical study 
on the massless and massive NL$x$
impact factor and found out that switching on a non-zero
quark mass has as effect to amplify the magnitude of the overall corrections
in the $\kd^2/m^2 < 1$ region, while keeping them positive.
We consider the re-calculated
finite part of the heavy quark impact factor 
presented here as the first step toward new LHC
phenomenological studies within the $k_T$-factorization scheme
of processes with heavy quarks in the final state.
Our next immediate project is to
study the cross section for
single bottom quark forward production at the LHC.

\section*{Acknowledgments}

This work has been supported by the 
Research Executive Agency (REA) of the European Union under 
the Grant Agreement number PITN-GA-2010-264564 (LHCPhenoNet),
by the Spanish Government and EU ERDF funds 
(grants FPA2007-60323, FPA2011-23778 and CSD2007-00042 
Consolider Project CPAN) and by GV (PROMETEUII/2013/007). 
GC acknowledges support from Marie Curie Actions (PIEF-GA-2011-298582).
MD acknowledges support from Juan de la Cierva programme (JCI-2011-11382).

\end{document}